\pdfoutput=1
\documentclass[conference]{IEEEtran}
\IEEEoverridecommandlockouts

\usepackage{cite}
\usepackage{amsmath,amssymb,amsfonts,amsthm}
\usepackage{graphicx}
\usepackage{textcomp}
\usepackage{xcolor}
\usepackage{booktabs}
\usepackage{tikz}
\usetikzlibrary{arrows.meta,positioning,calc}

\newtheorem{lemma}{Lemma}
\newtheorem{proposition}{Proposition}
\newtheorem{theorem}{Theorem}

\newcommand{\hd}[1]{\begin{tabular}[b]{@{}c@{}}#1\end{tabular}}

\begin{document}

\title{The Vulnerability of AFDM Chirp-Parameter Secrecy to Pilot-Based Estimation}

\author{
\IEEEauthorblockN{Mohammad Tasin and Muhammad Abdullah Arafat}
\IEEEauthorblockA{Department of Electrical and Electronic Engineering\\
Bangladesh University of Engineering and Technology, Dhaka, Bangladesh\\
Email: tasin2006072@gmail.com, abdullah\_arafat@eee.buet.ac.bd}
}

\maketitle

\begin{abstract}
It has been proposed to keep the chirp parameter $c_1$ of affine frequency division multiplexing (AFDM) secret and to measure physical-layer security by the cost of a brute-force search for it. We show that the channel-estimation pilot that AFDM transmits in every frame is itself a chirp at rate $c_1$ and leaks it to a passive eavesdropper. A lag-product search on the public grid followed by dechirp-FFT refinement recovers $c_1$ in 808 FFTs (0.33~s) without exhaustive search, and its root-mean-square error saturates at $2.8\times10^{-7}$, 4.4 times below the error that a receiver tolerates at 20~dB. In link-level simulations of a 4-path reference channel ($1.16$~Mbit per point), the eavesdropper's bit error rate (BER) stays within 15\% of that of the legitimate receiver, which estimates the channel from the pilot; with random delay-Doppler profiles, $c_1$ is still recovered to decoding precision in 88--94\% of the channel realizations. Securing the second chirp parameter $c_2$ does not prevent the leak. A per-frame, key-derived phase mask on the whole frame closes it: we prove that the mask makes the distribution of every observation independent of $c_1$, and in simulation the eavesdropper's BER is $0.50$ even when she knows $c_1$ exactly. Channel estimation under the mask costs the legitimate receiver 0.2--0.3~dB at BER from $10^{-1}$ to $10^{-3}$.
\end{abstract}


\section{Introduction}
Affine frequency division multiplexing (AFDM) is a chirp-based multicarrier waveform that achieves full diversity in doubly-dispersive channels and is a candidate for high-mobility next-generation networks~\cite{bemani2023afdm}. A receiver must know its two chirp parameters: $c_1$, set by the Doppler spread, and $c_2$, a free design parameter. Savaux \emph{et al.}~\cite{savaux2025robustness} therefore studied $c_1$ as a secret key against a passive eavesdropper who knows the channel: a brute-force search has complexity $\mathcal{O}(N^2)$, and wrong guesses leave the intercepted signal at a bit error rate (BER) near one half. Follow-up work secures $c_2$ instead. It hops the chirp parameters, including $c_2$, across symbols~\cite{chen2025hopping}, replaces $c_2m^2$ by a general phase function~\cite{liu2026phasefunction}, restricts $c_2$ to a secret codebook~\cite{wang2025seafdm,zhang2026codebook}, permutes the chirp sequence~\cite{rou2025chirppermuted}, or adds index modulation~\cite{cao2026isacafdm}; several of these works treat $c_1$ as public~\cite{liu2026phasefunction,zhang2026codebook,rou2025chirppermuted}. Pilot-phase encryption for AFDM on low-earth-orbit links~\cite{yuan2026leo} targets known-plaintext attacks on channel estimation, not the chirp rate, and delay-Doppler key generation for orthogonal time frequency space (OTFS) modulation~\cite{hadani2017otfs,li2026ddkeygen} derives keys from the channel.

None of these works asks what the pilot reveals. The pilot needed for channel estimation is a chirp at rate $c_1$ multiplied by a few path tones, and $c_2$ enters it only as a constant phase (Proposition~\ref{prop:c2_null}). Hence $c_1$ cannot be kept secret while the pilot is sent in the clear, whatever is done to $c_2$. This paper makes four contributions. (i) We show that classical chirp-rate estimation recovers $c_1$ from the pilot without exhaustive search, and compare it with the Cram\'er-Rao bound (CRB). (ii) We compare the estimation error with the error that a receiver tolerates, rather than with the key-grid spacing, and find it smaller, so a finer key grid cannot help. (iii) We evaluate the attack in a link-level simulator with a validated receiver, a data-bearing control, channel estimation at the legitimate receiver, random channel profiles and confidence intervals. (iv) We propose a per-frame keyed phase mask, prove that it makes the law of the received signal independent of $c_1$, and quantify its cost to the legitimate receiver, including channel estimation under the mask. Keyed time-domain randomization, with channel estimation that only the key holder can perform, has been proposed for OFDM~\cite{mohammed2019rand,mohammed2024enc}; here it is derived from, and analysed for, the leakage of the AFDM pilot.

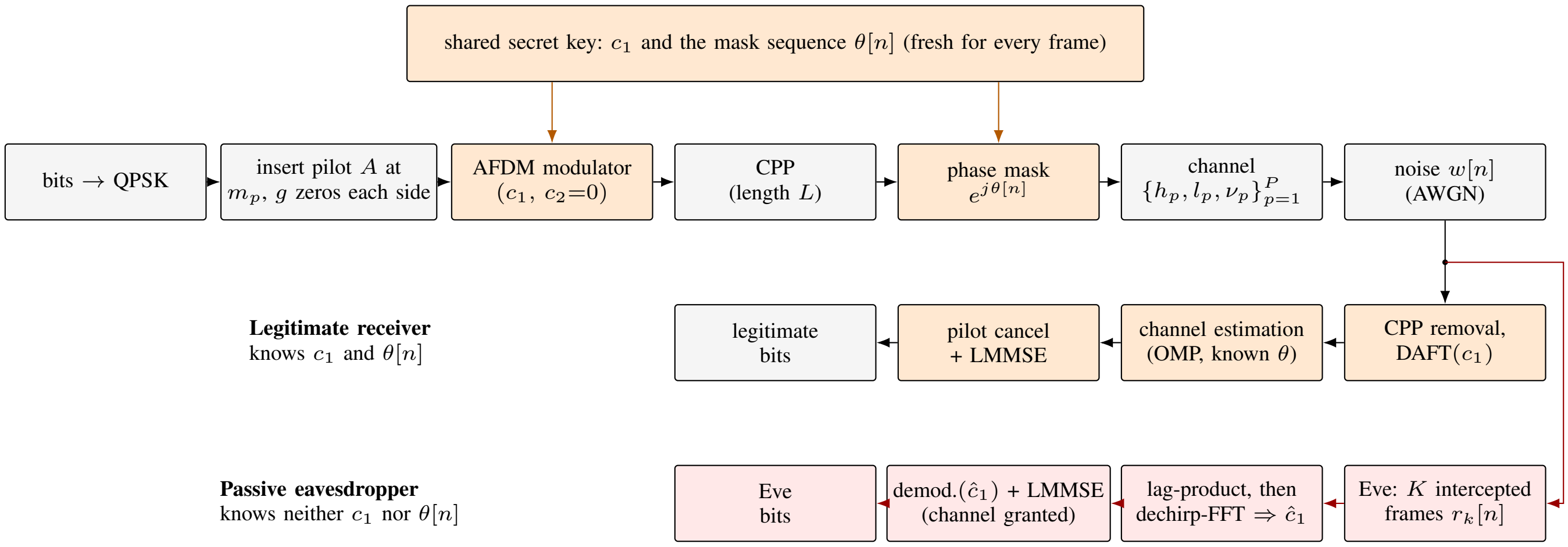
\begin{figure*}[t]
\centering
\begin{tikzpicture}[x=1in,y=1in,font=\scriptsize,
  blk/.style={draw, rounded corners=1pt, minimum height=0.34in, minimum width=0.9in, align=center, fill=gray!8, inner sep=2pt},
  key/.style={blk, fill=orange!18},
  eve/.style={blk, fill=red!9},
  ar/.style={-{Latex[length=1.5mm]}, line width=0.4pt}]
\node[blk] (bits) at (0.5,0) {bits $\to$ QPSK};
\node[blk] (ins) at (1.5,0) {insert pilot $A$ at\\$m_p$, $g$ zeros each side};
\node[key] (mod) at (2.5,0) {AFDM modulator\\$(c_1,\,c_2{=}0)$};
\node[blk] (cpp) at (3.5,0) {CPP\\(length $L$)};
\node[key] (mask) at (4.5,0) {phase mask\\$e^{j\theta[n]}$};
\node[blk] (chan) at (5.5,0) {channel\\$\{h_p,l_p,\nu_p\}_{p=1}^{P}$};
\node[blk] (noise) at (6.5,0) {noise $w[n]$\\(AWGN)};
\node[key, minimum width=3.3in] (kbox) at (3.5,0.62) {shared secret key: $c_1$ and the mask sequence $\theta[n]$ (fresh for every frame)};
\draw[ar, orange!70!black] (kbox.south -| mod) -- (mod.north);
\draw[ar, orange!70!black] (kbox.south -| mask) -- (mask.north);
\foreach \a/\b in {bits/ins, ins/mod, mod/cpp, cpp/mask, mask/chan, chan/noise} \draw[ar] (\a)--(\b);
\node[key] (cp) at (6.5,-0.72) {CPP removal,\\DAFT$(c_1)$};
\node[key] (est) at (5.5,-0.72) {channel estimation\\(OMP, known $\theta$)};
\node[key] (eq) at (4.5,-0.72) {pilot cancel\\+ LMMSE};
\node[blk] (out) at (3.5,-0.72) {legitimate\\bits};
\draw[ar] (noise.south) -- (cp.north);
\draw[ar] (cp)--(est); \draw[ar] (est)--(eq); \draw[ar] (eq)--(out);
\node[eve] (cap) at (6.5,-1.44) {Eve: $K$ intercepted\\frames $r_k[n]$};
\node[eve] (hat) at (5.5,-1.44) {lag-product, then\\dechirp-FFT $\Rightarrow\hat c_1$};
\node[eve] (edec) at (4.5,-1.44) {demod.$(\hat c_1)$ + LMMSE\\(channel granted)};
\node[eve] (eout) at (3.5,-1.44) {Eve\\bits};
\fill (6.5,-0.36) circle (0.9pt);
\draw[ar, red!60!black] (6.5,-0.36) -- (7.03,-0.36) -- (7.03,-1.44) -- (cap.east);
\draw[ar, red!60!black] (cap)--(hat); \draw[ar, red!60!black] (hat)--(edec); \draw[ar, red!60!black] (edec)--(eout);
\node[align=left] at (1.55,-0.72) {\textbf{Legitimate receiver}\\knows $c_1$ and $\theta[n]$};
\node[align=left] at (1.55,-1.44) {\textbf{Passive eavesdropper}\\knows neither $c_1$ nor $\theta[n]$};
\end{tikzpicture}
\caption{System and threat model; orange blocks use the key, red blocks belong to the eavesdropper.}
\label{fig:block}
\end{figure*}

\section{System Model}
Fig.~\ref{fig:block} shows the system and the threat model. Without the phase mask ($\theta[n]\equiv0$) it is the baseline AFDM pilot frame; the mask is the countermeasure of Sec.~\ref{sec:cm}.
\subsection{AFDM Waveform and Channel}
An AFDM frame carries $N$ symbols $x_m$ on chirp subcarriers,
\begin{equation}
s[n] = \tfrac{1}{\sqrt{N}}\textstyle\sum_{m=0}^{N-1} x_m e^{j2\pi(c_1 n^2 + c_2 m^2 + mn/N)},
\label{eq:afdm_signal}
\end{equation}
for $n=0,\dots,N-1$, i.e.\ $\mathbf{s}=\mathbf{\Lambda}_{c_1}\mathbf{F}_N^H\mathbf{\Lambda}_{c_2}\mathbf{x}$ with $\mathbf{F}_N$ the normalized DFT matrix and $[\mathbf{\Lambda}_c]_{n,n}=e^{j2\pi cn^2}$~\cite{bemani2023afdm}. A chirp-periodic prefix (CPP) of length $L\ge l_{\max}$ extends the frame to $M=N+L$ samples by $s[n]=s[N+n]e^{-j2\pi c_1(N^2+2Nn)}$, $n=-L,\dots,-1$, which continues the same chirp backwards in $n$. The frame crosses a doubly-dispersive channel with $P$ paths of complex gain $h_p$, integer delay $l_p\in[0,l_{\max}]$ and integer Doppler $\nu_p\in[-\nu_{\max},\nu_{\max}]$ (in bins of the discrete affine Fourier transform, DAFT),
\begin{equation}
r[n] = \textstyle\sum_{p=1}^{P} h_p\, s[n-l_p]\,e^{j2\pi \nu_p n/N} + w[n],
\label{eq:channel}
\end{equation}
with $w[n]\sim\mathcal{CN}(0,\sigma^2)$. The design value is $c_1=(2\nu_{\max}+1)/(2N)$. Since $c_1$ and $c_1+\tfrac12$ produce the same signal up to a cyclic shift of $N/2$ DAFT bins, $c_1$ is identifiable only modulo $\tfrac12$. We use $N=256$, $L=8$ ($M=264$), $P=4$, $l_{\max}=7$, $\nu_{\max}=3$ ($c_1=7/512$) and QPSK data. The reference channel has the delay-Doppler pairs $(l,\nu)\in\{(0,0),(2,1),(4,{-}2),(7,3)\}$ and gains $h_p\sim\mathcal{CN}(0,1/P)$, redrawn for every channel realization and constant over its frames; Sec.~\ref{sec:robust} also draws the pairs at random. The SNR is $E_s/N_0=1/\sigma^2$, with unit-energy data symbols, unit average channel energy and a unitary DAFT, so the noise has variance $\sigma^2$ per sample and per DAFT bin.

\subsection{Embedded Pilot and Legitimate Receiver}
\label{sec:rx}
Following the embedded-pilot design of~\cite{pilotafdm2022}, the symbol at $m_p=64$ is a pilot $x_{m_p}=A$ with $g$ zeros on each side. The pilot echoes fall on the DAFT bins $m_p+\nu_p-2Nc_1l_p$ ($2Nc_1=7$), so isolating them from the data requires $g=l_{\max}(2\nu_{\max}+1)+2\nu_{\max}=55$, which leaves $145$ data symbols per frame. The pilot has power $|A'|^2=|A|^2/N=3.98$ per time sample (6~dB above the power of a fully loaded frame), against $P_{\rm data}=145/256=0.566$ for the data, so it carries 88\% of the frame energy. The legitimate receiver (i) estimates $\{h_p,l_p,\nu_p\}$ by orthogonal matching pursuit (OMP) with $P$ atoms over the $56$ candidate pairs $(l,\nu)$, using the DAFT-domain response of the known pilot, (ii) subtracts the estimated pilot response, and (iii) detects the $145$ data symbols by linear minimum mean-square error (LMMSE) estimation with the estimated effective channel. Without step (ii), the strong pilot response enters the LMMSE estimates of the data and dominates the BER; the SNR definition keeps the pilot energy out of the SNR. Sanity checks against theory are in Sec.~\ref{sec:setup}. Superimposed pilots~\cite{pilotafdm2024ictt} are modulated by the same chirp and may leak as well; we do not study them.

\subsection{Threat Model}
Eve is passive and synchronized. To estimate $c_1$ she uses only the public frame structure: her estimator assumes a pilot made of a few tones, and Sec.~\ref{sec:robust} measures its sensitivity to that assumption. To decode she is granted, as in~\cite{savaux2025robustness}, the exact channel $\{h_p,l_p,\nu_p\}$ and $c_2$, and also the pilot amplitude, position and guard layout, so that the only missing quantity is $c_1$ (and the mask key in Sec.~\ref{sec:cm}). Her decoding results are therefore upper bounds on her performance. With $c_2$ known, we set $c_2=0$ without loss of generality; Proposition~\ref{prop:c2_null} shows that $c_1$ leaks for any $c_2$, known or secret.

\section{The Pilot Leak}
\begin{lemma}[Pilot chirp]
\label{lem:pilot_chirp}
The response of the channel \eqref{eq:channel} to the pilot alone ($x_m=A$ for $m=m_p$, $0$ otherwise) is a single chirp at rate $c_1$ times a sum of $P$ path tones,
\begin{equation}
r_A[n] = A' e^{j2\pi c_1 n^2} \textstyle\sum_{p=1}^{P} \alpha_p e^{j2\pi f_p n},
\label{eq:lemma1}
\end{equation}
with $A' = (A/\sqrt{N})e^{j2\pi c_2 m_p^2}$, $\alpha_p = h_p e^{j2\pi(c_1 l_p^2 - l_p m_p/N)}$ and $f_p = (m_p+\nu_p)/N - 2c_1 l_p$. The received signal is $r[n]=r_A[n]+r_D[n]+w[n]$, with $r_D$ the response to the data.
\end{lemma}
\begin{proof}[Proof sketch]
Expand $c_1(n-l_p)^2=c_1n^2-2c_1l_pn+c_1l_p^2$: the term $c_1n^2$ is common to all paths and factors out, $c_2m_p^2$ is constant, and the rest gives $\alpha_p$ and $f_p$. Because the CPP continues the chirp for $n<0$, \eqref{eq:lemma1} holds over the whole extended frame, except for the first $l_p$ samples of path $p$.
\end{proof}

\begin{proposition}[$c_2$-null]
\label{prop:c2_null}
$c_2$ enters \eqref{eq:lemma1} only through the constant $A'$, so knowing $c_2$ does not help to estimate $c_1$, and hiding $c_2$ does not prevent it. The same holds for any key-dependent phase or permutation applied in the DAFT domain~\cite{liu2026phasefunction,rou2025chirppermuted}: it changes $A'$, and at most the pilot position, which the estimator below does not use. An eavesdropper who keeps the prefix sees up to $M=N+L$ samples of the chirp.
\end{proposition}

\begin{theorem}[Cram\'er-Rao bound, dominant-tone model, cf.~\cite{peleg1991polyphase}]
\label{thm:crlb}
Let the pilot be one constant-modulus tone with unknown phase, frequency and chirp rate, in white Gaussian noise of variance $\sigma_{\rm eff}^2$, observed over $M$ consecutive samples in each of $K$ independent frames. Any unbiased $\hat c_1$ satisfies
\begin{equation}
\mathrm{var}(\hat{c}_1) \ge \frac{22.5}{\pi^2 \gamma M^5 K},\qquad \gamma=\frac{|A'|^2 G}{\sigma_{\rm eff}^2},\quad G=\textstyle\sum_p|h_p|^2.
\label{eq:crlb}
\end{equation}
\end{theorem}
\begin{proof}[Proof sketch]
The Fisher matrix of (phase, frequency, $2\pi c_1$) is $2\gamma\mathbf{H}$ with $\mathbf{H}$ the $3\times3$ Hankel matrix of the moments $\sum_n n^{i+k}$; $[\mathbf{H}^{-1}]_{33}\approx180/M^5$ for $M$ consecutive samples, so $\mathrm{var}(2\pi\hat c_1)\ge 90/(\gamma M^5 K)$.
\end{proof}
Data and pilot share the time samples, so we treat the data as Gaussian interference, $\sigma_{\rm eff}^2=\sigma^2+G P_{\rm data}$; $\gamma$, and with it the bound, then saturates at high SNR, at $\gamma=|A'|^2/P_{\rm data}=7.0$ (8.5~dB). For one tone, \eqref{eq:crlb} is the CRB up to the large-$M$ approximation. For the $P$-path pilot we computed the numerical CRB with unknown $c_1$ and an unknown complex gain and frequency for each path; it agrees with \eqref{eq:crlb} to within 1\% for the reference channel, whose tones are well separated. Because the data are not Gaussian and, after dechirping, the guard band keeps them away from the pilot tones, \eqref{eq:crlb} is a conservative reference for our estimator, not a strict bound.

\emph{Why the pilot leaks and the data do not.} By \eqref{eq:afdm_signal}, $s[n]=e^{j2\pi c_1n^2}u[n]$ over the whole frame, so the data also carry $c_1$, but only through the correlation of $u$. For i.i.d.\ unit-energy symbols on the set $\mathcal{D}$ of loaded bins, $\mathbb{E}\{s[n]s^*[n-\tau]\}=e^{j2\pi c_1(2n\tau-\tau^2)}\rho(\tau)$ with $\rho(\tau)=N^{-1}\sum_{m\in\mathcal{D}}e^{j2\pi m\tau/N}$, which is a tone at frequency $2c_1\tau$ in the lag product. A fully loaded frame has $\rho(\tau)=0$ for $0<\tau<N$ and leaks nothing at second order. The $111$ bins without data in our frame (pilot and guard) give $|\rho(\tau)|\le 0.31$ at the lags used below, enough to locate $c_1$ coarsely. The pilot adds a deterministic tone of amplitude $|A'|^2=3.98$, 13 times larger, and after dechirping it becomes $P$ strong tones, which precise estimation needs. Fig.~\ref{fig:leak} confirms the three cases (Sec.~\ref{sec:attack}).

\section{Attack and Results}
\label{sec:attack_res}
\subsection{Estimator and Setup}
\label{sec:setup}
Eve's estimator has two stages and uses neither $m_p$ nor $A$. (i) \emph{Coarse:} the spectra of the lag products $r_k[n]r_k^*[n-\tau]$, $\tau\in\{1,2,3,5\}$, summed over the $K$ intercepted frames and multiplied across lags, as in the product high-order ambiguity function~\cite{barbarossa1998haf}, are read at the frequencies $2c\tau$ of the $128$ public candidates $c=(2k+1)/(2N)$. (ii) \emph{Refinement:} four zoom stages of $201$ candidates (half-widths $3$, $0.1$, $0.006$ and $4\times10^{-4}$ grid units, one grid unit being $1/(2N)$) score each candidate $c$ by the sum of the $\hat P=4$ largest well-separated peaks of $|\mathrm{FFT}(\bar r[n]e^{-j2\pi cn^2})|^2$, where $\bar r=\sum_k r_k$ is the coherent sum of the frames over the $M$-sample window~\cite{peleg1991polyphase}; scoring only the strongest tone leaves a bias in multipath. This costs $4+4\times201=808$ FFTs, $0.33$~s on one CPU core irrespective of $K$. The coherent sum assumes a channel that is constant over the $K$ frames; $K=1$ needs no such assumption. Eve then decodes with the receiver of Sec.~\ref{sec:rx}, using $\hat c_1$ and the granted channel.

\begin{figure*}[t]
\centering
\includegraphics[width=\textwidth]{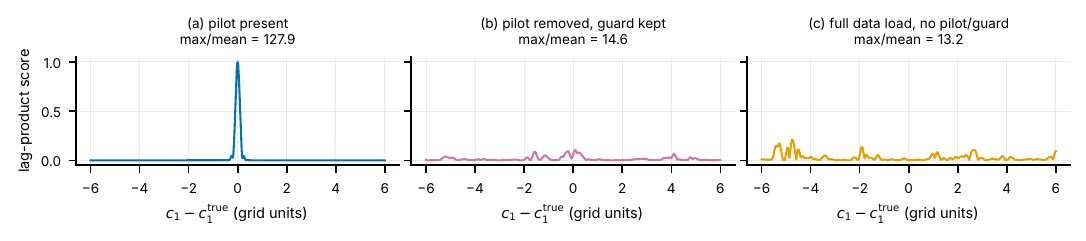}
\caption{Coarse lag-product score versus candidate $c_1$ for one channel realization ($K=8$ data-bearing frames, 0~dB).}
\label{fig:leak}
\end{figure*}
\begin{figure*}[t]
\centering
\includegraphics[width=\textwidth]{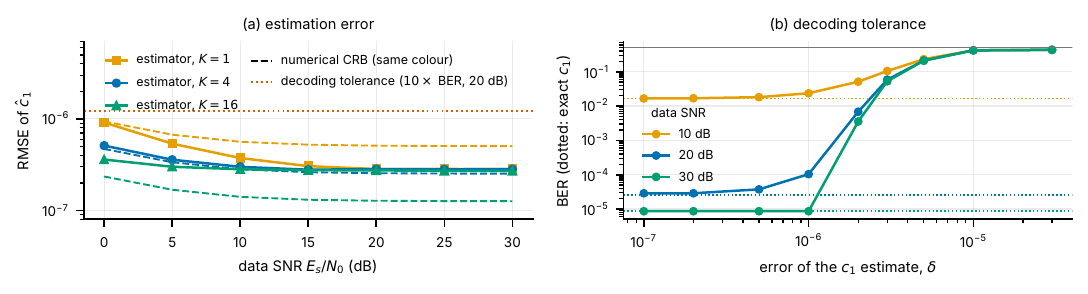}
\caption{(a) RMSE of $\hat c_1$ with the numerical CRB (dashed); (b) BER of a receiver with the exact channel versus the error $\delta$ in $c_1$.}
\label{fig:rmse}
\end{figure*}
\begin{figure}[t]
\centering
\includegraphics[width=\columnwidth]{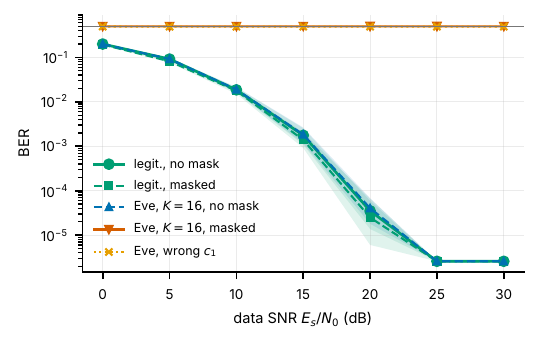}
\caption{BER versus SNR, reference channel (shading: 95\% CI).}
\label{fig:ber}
\end{figure}
Each BER point uses $250$ channel realizations $\times$ $16$ frames ($1\,160\,000$ data bits); 95\% confidence intervals are bootstrapped over realizations, and points with fewer than three errors are drawn at the rule-of-three level $2.6\times10^{-6}$. As a sanity check, the receiver reproduces the closed-form BER of QPSK in AWGN to within 1\% and in flat Rayleigh fading to within 10\% ($928$~kbit per point from 800 channel draws, for which one standard deviation of the Rayleigh estimate is 12\% at 15~dB and 21\% at 20~dB), and the $4$-path link beats flat Rayleigh, as full diversity requires.

\subsection{The Pilot Is the Leak}
\label{sec:attack}
Fig.~\ref{fig:leak} shows the coarse score for data-bearing frames in three cases. With the pilot (a), the score has one sharp peak, and over 100 trials per case the error $|\hat c_1-c_1|$ stays below the decoding tolerance $1.2\times10^{-6}$ of Sec.~\ref{sec:tol} in at least 0.98 of the trials for every $K\in\{1,4,16\}$ and SNR in $\{0,10,20\}$~dB, except for $K=1$ at 0~dB (0.86), where this 20-dB tolerance is conservative; above 0~dB one frame is enough. With only the pilot removed (b), at most 0.01 of the trials reach this precision, but the guard zeros still lead the coarse stage to the right grid cell in 0.14 of the trials at $K=1$ and 0~dB and in 1.00 at $K=16$ and $\ge10$~dB, as $\rho(\tau)\ne0$ predicts. A fully loaded frame (c) leaks neither (coarse success at most 0.02). The frame structure thus leaks the coarse value, and the pilot the precision needed to decode.

\subsection{Estimation Precision and Decoding Tolerance}
\label{sec:tol}
Fig.~\ref{fig:rmse}(a) shows the root-mean-square error (RMSE) of $\hat c_1$ over 150 trials per point, excluding gross errors ($|\hat c_1-c_1|>5\times10^{-5}$; none occurred). For $K=4$ the estimator is within 1.06--1.12 times the numerical CRB at every SNR from 0 to 30~dB. It is below the bound for $K=1$ (0.55--0.97 times), for the reason given after \eqref{eq:crlb}, and above it for $K=16$ (up to 2.1 times), because at every $K$ the RMSE saturates near $2.8\times10^{-7}$ above 15~dB. This floor is a bias of our multi-tone score under multipath, not a limit of the problem: it remains ($2.4\times10^{-7}$) with the data removed and the noise at $-60$~dB.

Security does not require Eve to reach the bound, only to stay below the precision a receiver needs. Fig.~\ref{fig:rmse}(b) ($348$~kbit per point, same rule-of-three convention; dotted: $\delta=0$) shows that a receiver with the exact channel and an error $\delta$ in $c_1$ keeps its BER within $2\times$ of the exact-$c_1$ BER for $\delta\le 0.66\times10^{-6}$ at 20~dB and within $10\times$ for $\delta\le 1.2\times10^{-6}$ ($4.2\times10^{-6}$ at 10~dB), and that its BER exceeds 0.4 for $\delta\ge10^{-5}$; this corresponds to the narrow mismatch window observed in~\cite{savaux2025robustness}. Eve's RMSE floor is 4.4 times below this $10\times$ tolerance at 20~dB (dotted line in Fig.~\ref{fig:rmse}(a)), and her median error at $K=16$ and 20~dB is $1.6\times10^{-7}$. The tolerance is a property of the waveform and is already $1\,603$ times finer than the grid spacing $1/(2N)=1.95\times10^{-3}$; since Eve does not search a grid, a finer key grid would not stop her. Below, $P_{\rm ok}$ is the fraction of trials with $|\hat c_1-c_1|<1.2\times10^{-6}$.

\subsection{BER of the Eavesdropper}
Fig.~\ref{fig:ber} and Table~\ref{tab:comp} give the link-level result for the reference channel; run times in Table~\ref{tab:comp} are for one CPU core. Without the mask, the legitimate receiver and Eve with $K=16$ frames are at $1.9\times10^{-2}$ and $1.9\times10^{-2}$ at 10~dB and at $3.5\times10^{-5}$ and $4.1\times10^{-5}$ at 20~dB, and their curves in Fig.~\ref{fig:ber} overlap. With $K=16$ her estimate is within the tolerance in every realization ($P_{\rm ok}=1$), and with $K=4$ in at least 0.99 of them. Eve's BER is never more than 15\% above the legitimate one, with $K=4$ as with $K=16$, and neither makes an error in $1.16$~Mbit at 25 and 30~dB. An Eve who guesses $c_1$ wrongly stays at $0.49$. Without the mask, channel estimation costs the legitimate receiver at most 0.03~dB at BER from $10^{-1}$ to $10^{-3}$.

\emph{Comparison with brute force and with existing protections.} A brute-force search with spacing $2\times1.2\times10^{-6}$ over $c_1\in(0,\tfrac12)$ never misses $c_1$ by more than the tolerance, so its BER is at most that of Fig.~\ref{fig:rmse}(b) at $\delta=1.2\times10^{-6}$. At 5.2~ms per hypothesis for a full decode test or 0.15~ms for one dechirp-FFT score, it costs 95 to 3\,210 times more than the attack. The rows marked $\ddagger$ in Table~\ref{tab:comp} apply the attack, in 150 trials per SNR, to the two kinds of protection that we can model, each with a fresh secret in every frame. A key-dependent $c_2$, which covers hiding, hopping and codebook selection of $c_2$~\cite{chen2025hopping,wang2025seafdm,zhang2026codebook}, leaves the median error at $2.3\times10^{-7}$, against $1.5\times10^{-7}$ without defence in the same trials, as Proposition~\ref{prop:c2_null} predicts. With $K=16$ a per-frame $c_2$ disturbs the coherent frame sum, so $P_{\rm ok}$ drops to 0.92--0.97, whereas a single frame is only rotated by a constant. Encrypting the pilot phase in every frame~\cite{yuan2026leo}, modelled as a random constant phase on the pilot symbol, leaves $P_{\rm ok}=1$ with $K=1$ (0.94--0.95 with $K=16$), since one frame suffices; for the same reason, hopping $c_1$ itself from frame to frame~\cite{chen2025hopping} would not hide it. Decoding the data under these schemes also needs the secret $c_2$ or pilot phase, so for them Table~\ref{tab:comp} reports only the leak of $c_1$; data secrecy when $c_2$ is secured is not studied here.

\begin{table*}[t]
\caption{Attack and defence on the reference channel ($N=256$).}
\label{tab:comp}
\centering
\footnotesize
\setlength{\tabcolsep}{4pt}
\begin{tabular}{@{}p{2.05in} p{1.55in} c c c c@{}}
\toprule
 & Cost per estimate of $c_1$ & \hd{$|\hat c_1-c_1|$,\\median, 20~dB} & \hd{$P_{\rm ok}$,\\10\,/\,20~dB} & \hd{BER,\\10~dB} & \hd{BER,\\20~dB}\\
\midrule
Legit., no mask & --- & --- & --- & $1.9\times10^{-2}$ & $3.5\times10^{-5}$\\
Eve, wrong $c_1$ (prior threat model) & --- & $\ge 5$ grid units & 0.00\,/\,0.00 & $0.49$ & $0.49$\\
Eve, brute force$^\dagger$ & $2.1\times10^{5}$ hypotheses, 31~s--18~min & $\le1.2\times10^{-6}$ & 1.00\,/\,1.00 & $\le2.8\times10^{-2}$ & $\le2.6\times10^{-4}$\\
Eve, pilot attack, $K{=}16$ & 808 FFTs, 0.33~s & $1.6\times10^{-7}$ & 1.00\,/\,1.00 & $1.9\times10^{-2}$ & $4.1\times10^{-5}$\\
\quad same, secret $c_2$ per frame$^\ddagger$ & 808 FFTs, 0.33~s & $2.3\times10^{-7}$ & 0.92\,/\,0.97 & --- & ---\\
\quad same, secret pilot phase per frame, $K{=}1^\ddagger$ & 808 FFTs, 0.33~s & $2.0\times10^{-7}$ & 1.00\,/\,1.00 & --- & ---\\
\midrule
Legit., masked & --- & --- & --- & $1.8\times10^{-2}$ & $2.5\times10^{-5}$\\
Legit., masked, exact channel (reference) & --- & --- & --- & $1.6\times10^{-2}$ & $1.1\times10^{-5}$\\
Eve, pilot attack, $K{=}16$, masked frames & 808 FFTs, 0.33~s & $81$ grid units & 0.00\,/\,0.00 & $0.50$ & $0.50$\\
Eve, exact $c_1$ but no key, masked frames & --- & $0$ & 1.00\,/\,1.00 & $0.50$ & $0.50$\\
\bottomrule
\end{tabular}

\vspace{2pt}
\footnotesize $^\dagger$ 31~s with a dechirp score, 18~min with a full decode test; BER bound: Fig.~\ref{fig:rmse}(b) at $\delta=1.2\times10^{-6}$.\quad $^\ddagger$ Decoding also needs the secret $c_2$ or pilot phase.
\end{table*}

\section{Countermeasure: Keyed Phase Mask}
\label{sec:cm}
\subsection{Design}
\label{sec:design}
The attack works because the pilot is a deterministic chirp over the whole $M$-sample window. We therefore multiply the whole CPP-extended frame by $e^{j\theta[n]}$, with $\theta[n]$ i.i.d.\ uniform on $[0,2\pi)$, fresh for each frame and generated from a key shared with the receiver by a stream cipher. This is a keyed phase scrambler: security comes from the key and the keystream, not from AFDM or $c_1$, and key exchange is outside the scope of this paper. The mask must cover the whole frame, because pilot and data are superimposed in time and the guard structure alone already leaks the coarse value (Fig.~\ref{fig:leak}(b)). The receiver cannot simply multiply $r[n]$ by $e^{-j\theta[n]}$: for $l_p\ne0$ this leaves the random factor $e^{j(\theta[n-l_p]-\theta[n])}$ on path $p$. Instead it folds the known mask into the effective channel, $\mathbf{H}_\theta=\sum_p h_p\mathbf{\Delta}^{\nu_p}\mathbf{\Pi}^{l_p}\mathbf{D}_\theta\mathbf{T}$, with $\mathbf{D}_\theta=\mathrm{diag}(e^{j\theta[n]})$, $\mathbf{T}$ the CPP-extended modulator, and $\mathbf{\Delta}$ and $\mathbf{\Pi}$ the Doppler and delay operators of \eqref{eq:channel}, at the same LMMSE cost as before. It also estimates the channel with the known mask: the masked pilot response of each of the $56$ candidate pairs $(l,\nu)$ is computed and fed to the same OMP.

\emph{Relation to OFDM and OTFS encryption.} Whether a keyed phase hides $c_1$ depends on where it acts. A phase per symbol in the DAFT domain, the AFDM counterpart of per-subcarrier phase encryption in OFDM and the form taken by the $c_2$, phase-function, permutation and pilot-phase schemes of Sec.~\ref{sec:attack_res}, only changes $A'$, and at most the pilot position, in \eqref{eq:lemma1} (Proposition~\ref{prop:c2_null}), so $c_1$ still leaks (Table~\ref{tab:comp}). A phase per time sample, as in our mask and in keyed time-domain randomization of OFDM~\cite{mohammed2019rand,mohammed2024enc}, randomizes every sample of the chirp; if it is i.i.d.\ uniform, Theorem~\ref{thm:inv} below applies. Delay-Doppler key generation for OTFS~\cite{li2026ddkeygen} produces keys rather than using them, and could supply the mask key. We did not simulate the OFDM and OTFS schemes.

\subsection{Why It Works}
\begin{theorem}[Invariance to $c_1$]
\label{thm:inv}
Let $\theta[n]$, $n=-L,\dots,N-1$, be i.i.d.\ uniform on $[0,2\pi)$, independent of the data, channel and noise, with independent masks in different frames. Then for every fixed data and channel, the joint law of the $K$ masked frames, and hence of every statistic computed from them, is the same for all values of $c_1$ (for any fixed $c_2$). The likelihood of the observation therefore does not depend on $c_1$, its Fisher information about $c_1$ is zero, and $I(c_1;\mathbf{r}\mid\mathbf{x},\mathbf{h})=0$.
\end{theorem}
\begin{proof}
By \eqref{eq:afdm_signal} and the CPP rule, $s[n]=e^{j2\pi c_1n^2}u[n]$ for $n=-L,\dots,N-1$, with $u[n]$ independent of $c_1$. The masked signal is $e^{j(\theta[n]+2\pi c_1n^2)}u[n]$. For fixed $c_1$, the map $\theta[n]\mapsto\theta[n]+2\pi c_1n^2 \pmod{2\pi}$ preserves the uniform law of each coordinate, so $\theta'[n]=\theta[n]+2\pi c_1n^2$ is again i.i.d.\ uniform and independent of $u$, channel and noise. Each received frame is a fixed function of $(u,\theta',\mathbf{h},\mathbf{w})$, whose law is thus independent of $c_1$, and the frames are independent given the data and channel.
\end{proof}
Theorem~\ref{thm:inv} does not assume a coherent eavesdropper: it holds for every function of the masked frames, whether it dechirps them, forms lag products or uses only their envelope, at any SNR and for any number of frames. For the coherent dechirp-FFT value $Z=\sum_n r[n]e^{-j2\pi(c_1n^2+f_pn)}$ at a pilot tone, the mask turns growth as $M$ into growth as $\sqrt M$, a power loss of $10\log_{10}M=24.2$~dB for a single tone without data or noise (19.6~dB measured at the strongest pilot tone of the reference channel at 10~dB, with data and noise). Table~\ref{tab:ks} tests the theorem with statistics of all three kinds at 10~dB, with the same channel and data and two values of $c_1$ $37$ grid units apart ($2000$ frames each): two-sample Kolmogorov--Smirnov (KS) tests reject equal distributions for all four statistics without the mask, and for none with it at the 5\% level.

\begin{table}[t]
\caption{KS tests of Theorem~\ref{thm:inv} at 10~dB.}
\label{tab:ks}
\centering
\footnotesize
\setlength{\tabcolsep}{3.5pt}
\begin{tabular}{@{}ll cc cc@{}}
\toprule
 & & \multicolumn{2}{c}{no mask} & \multicolumn{2}{c}{masked}\\
\cmidrule(lr){3-4}\cmidrule(lr){5-6}
Statistic & Kind & KS & $p$ & KS & $p$\\
\midrule
Dechirp-FFT peak & coherent & 1.00 & $<10^{-10}$ & 0.018 & 0.92\\
Lag-1 product peak & lag product & 1.00 & $<10^{-10}$ & 0.033 & 0.23\\
Pilot-tone bin & coherent & 1.00 & $<10^{-10}$ & 0.030 & 0.33\\
Envelope kurtosis & magnitude only & 0.75 & $<10^{-10}$ & 0.037 & 0.13\\
\bottomrule
\end{tabular}
\end{table}

The theorem assumes a truly random mask. With a stream cipher the guarantee becomes computational and rests on the keystream; we do not analyse keystream recovery by known-plaintext attacks or active attackers. It concerns $c_1$ only.

\subsection{Results}
With the mask, Eve's estimate no longer follows $c_1$: its median error is 80--81 grid units at every SNR ($P_{\rm ok}=0$), and her BER is $0.50$ at every SNR (Fig.~\ref{fig:ber}), as is the BER of an Eve who knows $c_1$ exactly but not the mask (Table~\ref{tab:comp}).

The legitimate receiver is never given the channel: it sees only the received samples, the pilot and the key, and estimates the channel as in Sec.~\ref{sec:design}. With the mask it is at $1.8\times10^{-2}$ at 10~dB and $2.5\times10^{-5}$ at 20~dB, against $1.6\times10^{-2}$ and $1.1\times10^{-5}$ for the exact-channel reference in Table~\ref{tab:comp} (given the true channel, for comparison only). Channel estimation under the mask thus costs 0.2--0.3~dB at BER from $10^{-1}$ to $10^{-3}$, and more at lower BER, because the mask spreads data energy into the guard window, where the estimator sees it as interference: the normalized mean-square error (NMSE) of the channel estimate saturates at $-27$~dB, against $-43$~dB at 20~dB and $-53$~dB at 30~dB without the mask. The masked link even needs 0.2 to 0.6~dB \emph{less} SNR than the unmasked one at BER from $10^{-1}$ to $10^{-3}$, plausibly because the detector also collects the data energy spread into the guard bins; we do not count this as a benefit.

\begin{table}[t]
\caption{Random delay-Doppler pairs; Eve with $K=16$.}
\label{tab:rand}
\centering
\footnotesize
\setlength{\tabcolsep}{3pt}
\begin{tabular}{@{}c cc c cc c@{}}
\toprule
 & \multicolumn{3}{c}{no mask} & \multicolumn{3}{c}{masked}\\
\cmidrule(lr){2-4}\cmidrule(lr){5-7}
SNR & legit. & Eve & $P_{\rm ok}$ & legit. & Eve & NMSE\\
\midrule
5~dB & $8.8\times10^{-2}$ & $9.5\times10^{-2}$ & 0.94 & $8.3\times10^{-2}$ & $0.50$ & $-25$~dB\\
15~dB & $4.1\times10^{-3}$ & $5.0\times10^{-3}$ & 0.94 & $1.4\times10^{-3}$ & $0.50$ & $-28$~dB\\
25~dB & $0$ & $4.8\times10^{-3}$ & 0.88 & $0$ & $0.50$ & $-28$~dB\\
\bottomrule
\end{tabular}
\end{table}

\section{Robustness}
\label{sec:robust}
\emph{Random channels.} Table~\ref{tab:rand} repeats the link simulation with distinct delay-Doppler pairs drawn at random in every realization ($80$ realizations $\times$ $16$ frames per SNR, $371$~kbit per point; NMSE: legitimate channel estimate under the mask; 0: no error). Eve with $K=16$ recovers $c_1$ to within the tolerance in 0.88--0.94 of the realizations ($P_{\rm ok}$), and in 100 trials per point at 5, 15 and 25~dB her RMSE is 5.7--$6.9\times10^{-7}$ for $K=4$ and $16$, up to 4.9 times the CRB. Her BER is 7\% and 22\% above the legitimate one at 5 and 15~dB, and at 25~dB the few failing realizations leave her at $4.8\times10^{-3}$ while the legitimate receiver makes no error. With the mask she is at $0.50$, and the legitimate channel NMSE is $-25$ to $-28$~dB.

\emph{Continuous $c_1$.} Savaux \emph{et al.}~\cite{savaux2025robustness} search $c_1$ over a continuous range. With $c_1$ drawn uniformly within $\pm\tfrac12$ grid unit of a grid point, a fraction 0.06--0.10 of the trials at $K=4$ and 0--20~dB are gross errors, and the RMSE of the rest is 1.02--1.12 times the CRB (100 trials per point).

\emph{Assumed number of tones.} At $K=4$ and 10~dB, assuming $\hat P\in\{3,4,5\}$ tones for the true $P=4$ keeps the RMSE within 1.00--1.06 times the CRB, and $\hat P\in\{2,3,4,5,6,8\}$ within 1.41 times, without gross errors (100 trials each).

\emph{Limits.} The run times depend on the implementation, and channels with more paths, closer tones or fractional delays and Dopplers may change the RMSE floor of Fig.~\ref{fig:rmse}(a).

\section{Conclusion}
The AFDM channel-estimation pilot is a chirp at rate $c_1$, so a passive eavesdropper recovers the supposedly secret parameter by classical chirp-rate estimation in 808 FFTs, below the precision needed to decode, whatever is done to $c_2$. This breaks the brute-force threat model of~\cite{savaux2025robustness}, in which $c_1$ is the secret, and shows that securing $c_2$~\cite{chen2025hopping,liu2026phasefunction,wang2025seafdm,zhang2026codebook,rou2025chirppermuted} does not hide $c_1$: without a mask such as ours, $c_1$ must be treated as public, as~\cite{liu2026phasefunction,zhang2026codebook,rou2025chirppermuted} already assume. A per-frame keyed phase mask makes every observation independent of $c_1$; in simulation it takes the eavesdropper to chance, and channel estimation under the mask costs the legitimate receiver 0.2--0.3~dB at BER from $10^{-1}$ to $10^{-3}$. Open points are the security of the keystream against known-plaintext attacks, an eavesdropper who must estimate the channel herself, data secrecy when $c_2$ is secured, fractional delays and Dopplers, and other chirp-based waveforms with a similar pilot.

\end{document}